\documentclass[apj]{emulateapj}
\usepackage[]{natbib}

\received{26 June 2008}
\accepted{3 October 2008}
\slugcomment{{\it The Astrophysical Journal}}
\shorttitle{On the Origin of \tev}
\shortauthors{Mukherjee et al.}

\begin{document}
\def\ro{{\it ROSAT\/}}
\def\xmm{{\it XMM-Newton}}
\def\chandra{{\it Chandra}}
\def\swift{{\it Swift}}

\def\apsr{PSR J1833$-$0827}
\def\tev{HESS~J1834$-$087}
\def\hess{{\it HESS}}
\def\psr{XMMU~J183435.3$-$084443}
\def\pwn{G23.234$-$0.317}
\def\snr{G23.3$-$0.3}
\def\etal{et al.}

\title{On the Origin of TeV Gamma-ray Emission from \tev}

\author {R. Mukherjee\altaffilmark{1}, E. V. Gotthelf\altaffilmark{2}, and J. P. Halpern\altaffilmark{2}}
\altaffiltext{1}
{Department of Physics \& Astronomy, Barnard College, Columbia University,
New York, NY 10027}
\altaffiltext{2}
{Columbia Astrophysics Laboratory, Columbia University, New York, NY 10027}

\begin{abstract}
We present an X-ray study of the field containing the extended TeV
source \tev\ using data obtained with the \xmm\ telescope.  Previously,
the coincidence of this source with both the shell-type supernova remnant
(SNR) W41 and a giant molecular cloud (GMC) was interpreted as
favoring $\pi^0$-decay $\gamma$-rays from interaction of the old SNR with the
GMC. Alternatively, the TeV emission has been attributed to inverse Compton
scattering from leptons deposited by \apsr, a pulsar assumed to
have been born in W41
but now located $24^{\prime}$ from the center of the SNR (and the TeV source).
Instead, we argue for a third possibility, that the TeV emission is
powered by a previously unknown pulsar wind nebula located near the center of W41.
The candidate pulsar is \psr, a hard X-ray point source that lacks an optical
counterpart to $R>21$ and is coincident with diffuse X-ray emission.
The X-rays from both the point source and diffuse feature are
evidently non-thermal and highly absorbed. A best fit power-law model
yields photon index $\Gamma \sim 0.2$ and $\Gamma \sim 1.9$, for the
point source and diffuse emission, respectively, and $2-10$~keV flux
$\approx 5 \times 10^{-13}$~ ergs~cm$^{-2}$~s$^{-1}$ for each. At the
measured 4~kpc distance of W41, the observed X-ray luminosity implies an energetic
pulsar with $\dot E \sim 10^{36}\ d_4^2$~ergs~s$^{-1}$, which is
also sufficient to generate the observed $\gamma$-ray luminosity
of $2.7 \times 10^{34}\ d_4^2$~ergs~s$^{-1}$ via inverse Compton scattering.

\end{abstract}


\section{Introduction}

The HESS atmospheric Cherenkov telescope system has surveyed the
Galactic plane with unprecedented sensitivity and spatial resolution,
revealing $\approx 50$ previously unknown sources of TeV
($>10^{11}$~eV) $\gamma$-ray emission \citep{aha06a,aha08a}.  About
half of the Galactic TeV sources detected by HESS are identified with
supernova products\footnote{\url{http://www.mpi-hd.mpg.de/hfm/HESS/public/HESS\_catalog.htm}}
-- supernova remnants (SNRs) or pulsar wind nebulae (PWNe), which is a
likely scenario for many of the unclassified sources as well.  For
PWNe, the leading explanation for the generation of TeV emission is
leptonic, inverse Compton scattering of the microwave background or
possibly ambient IR photons \citep{aha06b}.  TeV $\gamma$-rays can
also arise from the decay of neutral pions created in hadronic
collisions of high-energy protons with the ambient medium.

The origin of \tev\ is a matter of great interest. This extended
source ($5.4^\prime$ radius at 1 $\sigma$) lies close to the center of
the large ($27^\prime$ diameter) shell-type SNR W41 (\snr) and has a
$\gamma$-ray flux of 8\% that of the Crab above 200 GeV
\citep{aha06a}.  The TeV flux and extended nature of \tev\ were
confirmed with the MAGIC $\gamma$-ray telescope \citep{alb06}.  Based
on VLA Galactic plane survey data, CO molecular line data, and an
X-ray observation, \citet{tia07} argue that this TeV source could be
$\pi^0$-decay $\gamma$-rays from the interaction of the old ($\sim
10^{5}$~yr) SNR W41 with a giant molecular cloud (GMC), as envisioned
by \citet{yam06}.  This would be an important result showing that old
SNR shells are potential sources of cosmic rays.  Alternatively the
location of the pulsar \apsr, $24^{\prime}$ from the center of \tev,
prompted \citet{bar08} to favor it as a source of relativistic
electrons, injected earlier at the birthplace of the pulsar, to power
the TeV emission.

Herein, we present further X-ray analysis of the \xmm\ observation of
\tev.  We detect a point source and associated diffuse emission at the
center of W41, which are likely a pulsar and wind nebula.  This is a
hard, non-thermal point source in the \xmm\ image, and it lacks an
optical/IR counterpart. Based on the morphological and spectral
evidence, we argue that this point source is a pulsar that could power
the observed TeV emission from \tev, as in other PWNe.

\section{Data Analysis and Results} 

The field containing \tev\ was observed by \xmm\ for 20 ks on 2005
September (P.I. P\"uhlhofer, ObsID 0302560301).  \citet{tia07} have
already analyzed the data from the pn~detector of the European Photon
Imaging Camera (EPIC).  Here, we complete the study of this
observation by including data from the two EPIC MOS CCD cameras (MOS1
and MOS2). These instruments were operated in ``full frame'' mode
using the medium filter with \tev\ located at the center of the
$0.5^{\circ}$ diameter field-of-view. This mode provides a time
resolution of 2.6~s, insufficient to search for a signal from a
typical rotation-powered pulsar.  We analyzed standard data products
obtained from the \xmm\ archive (processed with the
xmmsas\_20050428\_1800-6.0.0 pipeline) using both the SAS and FTOOLS
software. The observation was contaminated by several intervals of
high particle background, which were filtered out to leave a total of
13.86~ks of good observing time per detector.  Details of the EPIC~pn
observation can be found in \citet{tia07}.

\begin{figure}[t!]
\centerline{
\includegraphics[width=0.95\linewidth]{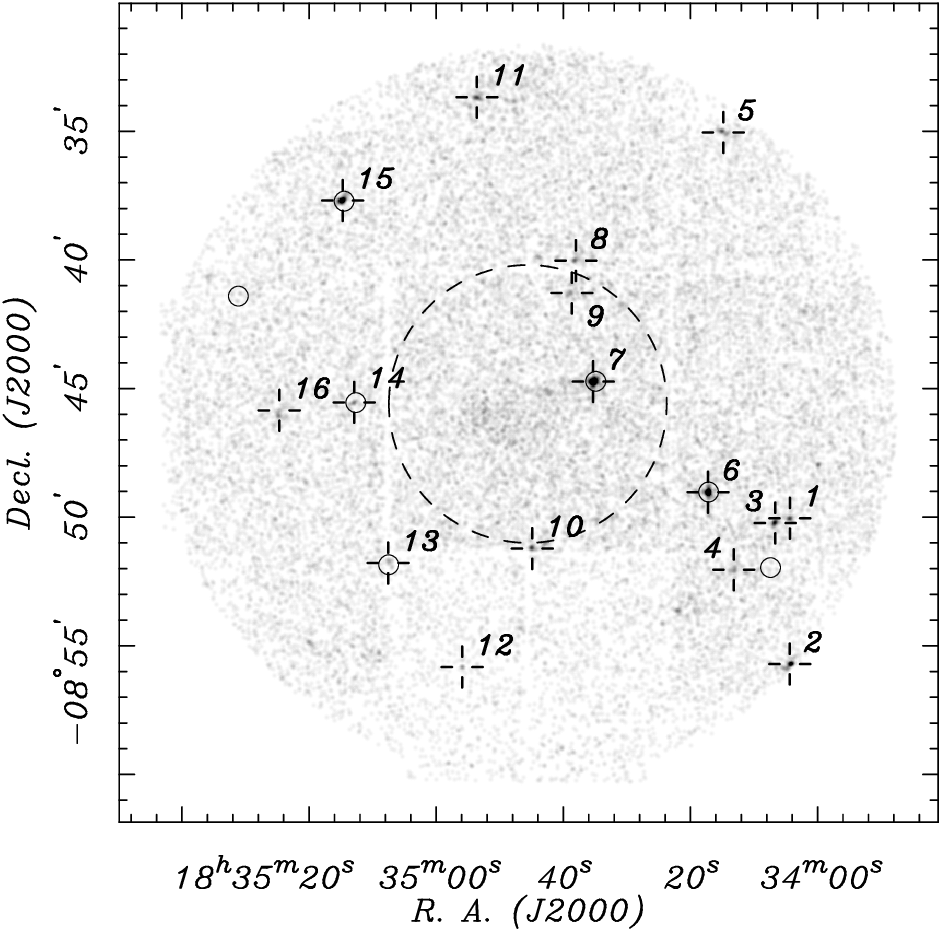}
}
\caption{\xmm\ MOS exposure corrected image of the \tev\ field in the
  1.5--7~keV energy band, selected to best reveal the central diffuse
  emission.  The dashed circle is the $1\ \sigma$ extent of the TeV
  source.  Significant \xmm\ point sources are numbered; their
  properties are listed in Table 1. \psr\ (source 7), the hardest
  source in the \xmm\ data, falls within the central $\gamma$-ray
  emission.  The seven small circles mark the locations of
  sources detected in \swift\ XRT images.}
\label{mosimg}
\end{figure}

\begin{figure*}
\centerline{
\includegraphics[width=0.46\textwidth]{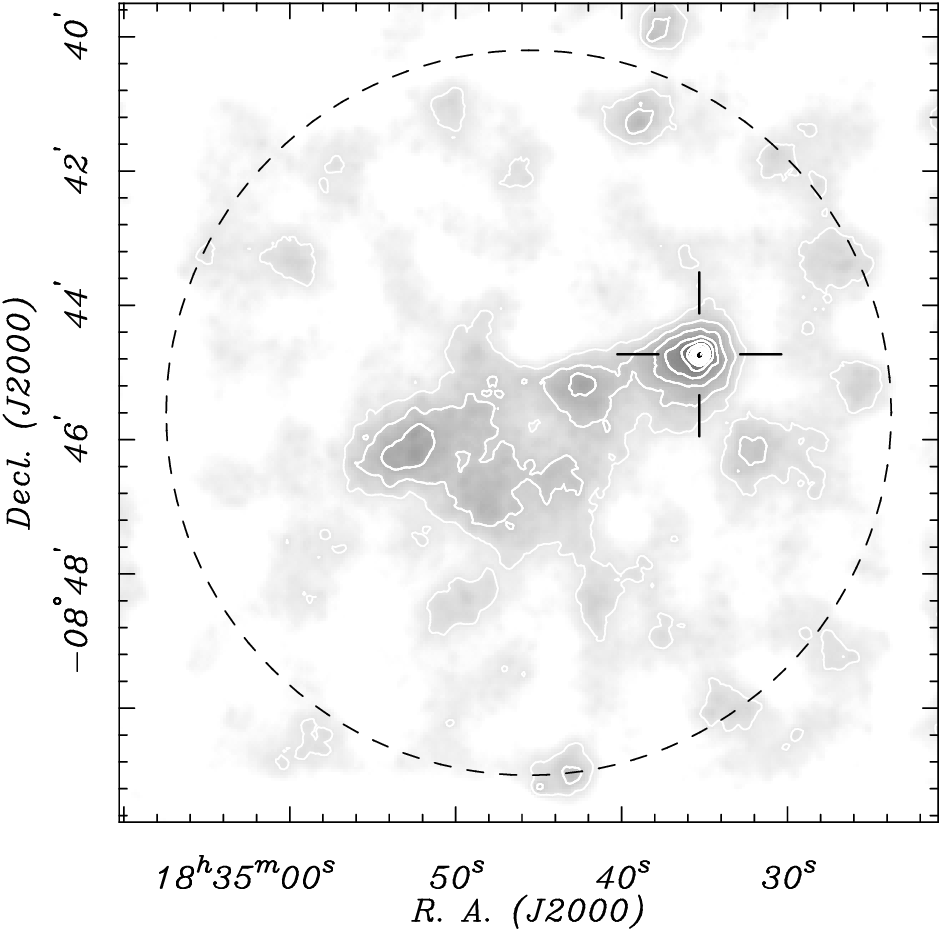}
\hfil
\includegraphics[width=0.46\textwidth]{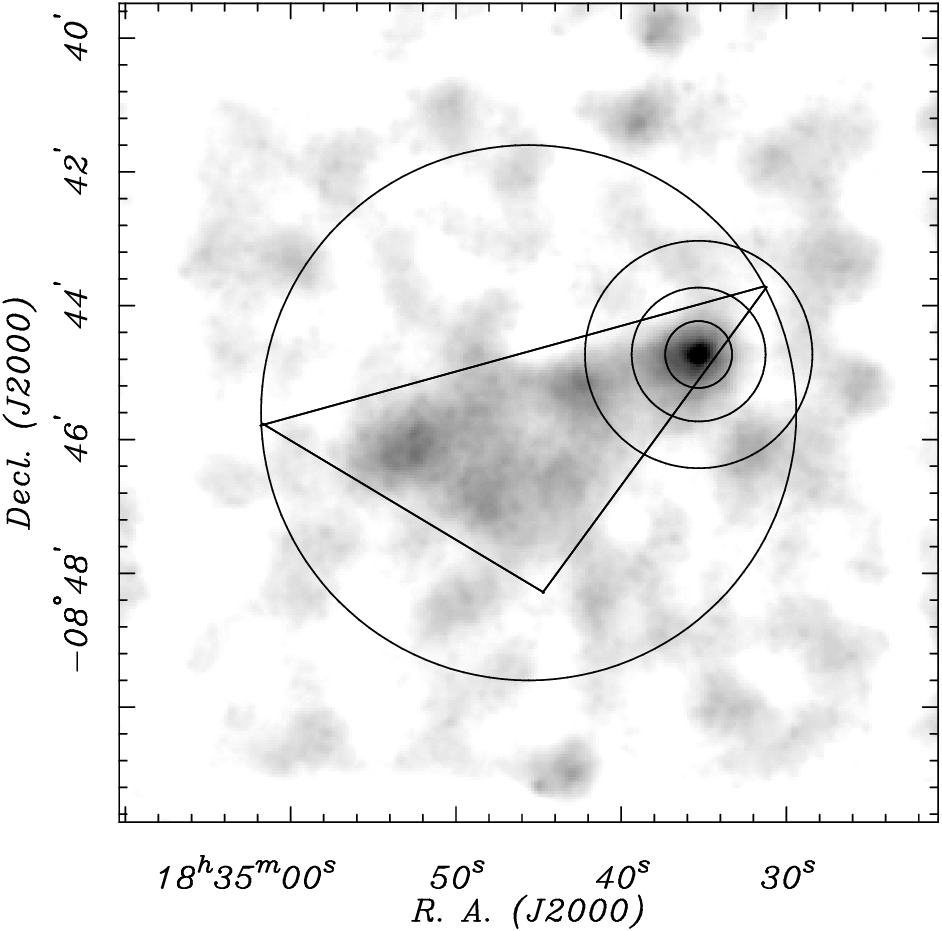}
}
\caption{A zoom-in on the \tev\ region shown in Figure~1. {\it Left\/}:
Image is smoothed and scaled to highlight the central diffuse emission and its
connection to point source 7 marked by the cross. The
greyscale intensity is linear and the contours are uniformly spaced.
The dashed circle reproduces the $1\ \sigma$ extent of the TeV
source. {\it Right\/}: Outlines of the spectral extraction
and background regions described in the text.}
\label{images}
\end{figure*}

\begin{deluxetable*}{lccrrrcccccccc}
\tablecolumns{14}
\tablewidth{0pt}
\tablecaption{Point Sources in \xmm\ ObsID 0302560301 and Nearest Optical/IR Candidates}
\tablehead
{
\colhead{No.} & \multicolumn{2}{c}{X-ray Position (J2000.0)} & \colhead{$N_s$\tablenotemark{a}} & \colhead{$N_h$\tablenotemark{a}} &
 & \multicolumn{2}{c}{Opt./IR Position (J2000.0)} & \colhead{$B2$\tablenotemark{b}} & \colhead{$R2$\tablenotemark{b}}
 & \colhead{$J$\tablenotemark{c}} & \colhead{$H$\tablenotemark{c}} & \colhead{$K$\tablenotemark{c}} & \colhead{Offset}\\
 & \colhead{R.A.} & \colhead{Decl.} & \colhead{(cts)} & \colhead{(cts)} & \colhead{HR\tablenotemark{a}}
 & \colhead{R.A.} & \colhead{Decl.} & \colhead{(mag)} & \colhead{(mag)} & \colhead{(mag)}& \colhead{(mag)} & \colhead{(mag)}
 & ($^{\prime\prime}$)
}
\startdata
 1\dotfill & 18 34 04.38 & $-$08 50 02.7  & 121 &  15 & $-$0.78 &  \dots      &  \dots        & \dots & \dots & \dots & \dots & \dots & \dots \\
 2\dotfill & 18 34 04.38 & $-$08 55 42.6  &   1 &  67 &    0.97 & 18 34 04.42 & $-$08 55 42.4 & 20.89 & 18.93 & 16.04 & 15.17 & 13.47 &  0.6  \\
 3\dotfill & 18 34 06.67 & $-$08 50 13.8  &  18 & 110 &    0.71 & 18 34 06.67 & $-$08 50 15.1 & \dots & \dots & \dots & 14.60 & 12.98 &  1.4  \\
 4\dotfill & 18 34 13.20 & $-$08 52 02.7  &  29 &  25 & $-$0.08 &  \dots      &  \dots        & \dots & \dots & \dots & \dots & \dots & \dots \\
 5\dotfill & 18 34 14.85 & $-$08 35 02.8  &  96 &  56 & $-$0.26 & 18 34 14.70 & $-$08 35 01.1 & 17.13 & 14.87 & 12.27 & 11.59 & 11.37 &  2.9  \\
 6\dotfill & 18 34 17.24 & $-$08 49 01.8  &  66 & 140 &    0.36 & 18 34 17.28 & $-$08 49 01.8 & \dots & \dots & \dots & \dots & \dots & \dots \\
 7\dotfill & 18 34 35.32 & $-$08 44 43.8  &   0 & 258 &    1.00 &  \dots      &  \dots        & \dots & \dots & \dots & \dots & \dots & \dots \\
 8\dotfill & 18 34 38.02 & $-$08 40 01.9  &  40 &  11 & $-$0.58 &  \dots      &  \dots        & \dots & \dots & \dots & \dots & \dots & \dots \\
 9\dotfill & 18 34 38.67 & $-$08 41 17.3  &  38 &  55 &    0.19 & 18 34 38.75 & $-$08 41 19.2 & \dots & \dots & 16.30 & 15.23 & \dots &  2.3  \\
10\dotfill & 18 34 44.90 & $-$08 51 13.3  &   2 &  80 &    0.96 & 18 34 45.00 & $-$08 51 10.3 & 19.98 & 16.39 & \dots & \dots & \dots &  0.9  \\
11\dotfill & 18 34 53.61 & $-$08 33 40.6  &  30 &  18 & $-$0.26 &  \dots      &  \dots        & \dots & \dots & \dots & \dots & \dots & \dots \\
12\dotfill & 18 34 55.93 & $-$08 55 49.6  &   9 &  16 &    0.25 &  \dots      &  \dots        & \dots & \dots & \dots & \dots & \dots & \dots \\
13\dotfill & 18 35 07.57 & $-$08 51 46.9  &  44 &  51 &    0.07 &  \dots      &  \dots        & \dots & \dots & \dots & \dots & \dots & \dots \\
14\dotfill & 18 35 12.91 & $-$08 45 34.6  &   0 &  57 &    1.00 &  \dots      &  \dots        & \dots & \dots & \dots & \dots & \dots & \dots \\
15\dotfill & 18 35 14.71 & $-$08 37 41.5  & 158 & 147 & $-$0.04 & 18 35 14.66 & $-$08 37 40.6 & \dots & \dots & 14.71 & 11.76 & 10.42 &  1.3  \\ 
16\dotfill & 18 35 24.70 & $-$08 45 51.0  &  35 &  59 &    0.26 &  \dots      &  \dots        & \dots & \dots & \dots & \dots & \dots & \dots \\
\enddata
\tablecomments{Units of right ascension are hours, minutes, and seconds, and units of declination are degrees, arcminutes, and arcseconds.}
\tablenotetext{a}{Background subtracted counts using data from all EPIC detectors.  Hardness ratio defined as HR=$(N_h-N_s)/(N_h+N_s)$,
where $N_s$ and $N_h$ are the counts  in the 0.3$-2$ keV and 2$-$10 keV bands, respectively.}
\tablenotetext{b}{Magnitude from the USNO-B1.0 Catalog.}
\tablenotetext{c}{Magnitude from the 2MASS Catalog.}
\label{srctab}
\end{deluxetable*}

We constructed exposure corrected, merged images from the two MOS
cameras in various X-ray energy bands.  Figure~\ref{mosimg} shows the
full \xmm\ MOS 1.5$-$7~keV X-ray image of the \tev\ field with the
significant ($>4\ \sigma$) point sources marked. The dashed circle
indicates the 1 $\sigma$ extent of the TeV emission \citep{aha06a}.
Table~\ref{srctab} lists the basic X-ray data on the point sources,
and the most likely USNO-B1.0 and/or 2MASS counterparts of some of
them.  \psr\ (source~7) is the hardest one. In the EPIC-pn detector,
source~7 fell on a gap between the CCDs, so it appears as two faint
point sources in the image analysis of \citet[][Fig.~4]{tia07}.  It is
also detected by the \swift\ X-ray telescope (XRT) in an observation of the
\tev\ field by \citet{lan06} (see \S2.1).  

Figure~\ref{images} is a zoom-in on the region near source~7,
showing the diffuse X-ray emission, hereafter referred to as \pwn,
extending from the point source throughout much of the central part of
the TeV emission.  Compared to the image analysis of the EPIC~pn data
by \citet{tia07}, the extent of the diffuse X-ray emission is
significantly larger and clearly merges with the hard point source.
It lies near the center of the SNR, in a region of enhanced radio
emission.

We extracted spectra for source~7 and \pwn\ using data from each EPIC
instrument and fitted them to a power-law model using the XSPEC
software. For the point source we used a radius $r=0.5'$ circular
aperture and a concentric $1^{\prime}<r \leq 1.7^{\prime}$ annular
background region.  The diffuse emission is extracted from a
triangular region (see Fig.~\ref{images}) with the point source
region excluded; the background is a $r=4^{\prime}$ circle with this diffuse
region punched out. Spectra from the two MOS cameras were
summed. Fitting of these spectra is limited by the small number of
source counts for each region per EPIC instrument, which were grouped
with a minimum of 40 or 20 counts per spectral channel for the pn and
MOS point source data, respectively, and 150 counts per channel for
the diffuse emission. The pn counts for the point source are 2/3 of
the summed MOS counts because most of the pn counts are lost in the
gap between the CCDs.  Spectra from both regions are evidently highly
absorbed as the flux below 2~keV is strongly cut off.

\begin{figure}[t]
\centerline{
\includegraphics[height=0.95\linewidth,angle=270,clip=true]{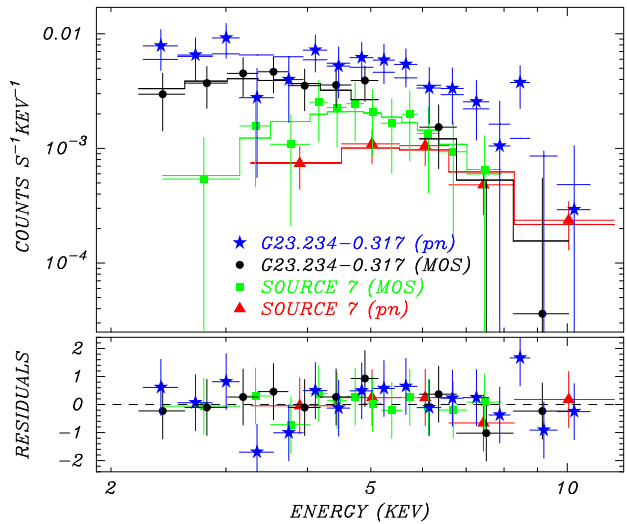}
}
\caption{\xmm\ spectra of source 7 in the MOS ({\it squares})
and pn ({\it triangles}), and the diffuse source \pwn\ (MOS: {\it circles},
pn: {\it stars}) fitted
to the power-law model described in the text. Residuals from this fit
are displayed in the lower panel.}
\label{spectra}
\end{figure}

Figure~\ref{spectra} shows the pairs of spectra from each region and
instrument fitted to an absorbed power law with column density $N_{\rm
H}$ and photon index $\Gamma$ tied for each extraction region.
Although we cannot distinguish between continuum spectral models, we
see no evidence of line emission in the background subtracted
spectra. Table~\ref{spectab} presents the best-fit spectral parameters
along with the measured flux, taken as the average of the EPIC MOS and
pn. Although the error on each parameter is large, the estimated
2$-$10~keV fluxes are fairly robust for the range of allowed
values. Although the X-ray measured column density for source~7 and
\pwn\ is not well constrained, it likely exceeds the integrated Galactic
total\footnote{Galactic $N_{\rm H}$ calculator available at
\url{heasarc.gsfc.nasa.gov/cgi-bin/Tools/w3nh/w3nh.pl}} in this
direction of $N_{\rm H} = 2 \times 10^{22}$~cm$^{-2}$, suggesting that
molecular material may lie in front of this structure.  In
Table~\ref{spectab} we also present spectral fits allowing independent
values for the column density of each region.

\begin{deluxetable}{lcc}
\tablecolumns{3}
\tablewidth{0pt}
\tablecaption{\xmm\ Spectroscopy}
\tablehead
{
Parameter  & Source 7 & \pwn\
}
\startdata
$N_{\rm H}$ (cm$^{-2}$)          & $6.0^{+2.6}_{-5.3} \times 10^{22}$ &  tied\\
 $\Gamma$                        & $0.2^{+0.9}_{-1.0}$                 & $1.9^{+1.0}_{-0.9}$  \\
$F_x$ (ergs~s$^{-1}$~cm$^{-2}$)\tablenotemark{a} & $4.9\times10^{-13}$                & $4.0\times10^{-13}$\\
$\chi^2$~(dof)                   & \multicolumn{2}{c}{ 0.64(20) } \\
\hline
 $N_{\rm H}$ (cm$^{-2}$)         & $13^{+31}_{-12}\times 10^{22}$     & $3.3^{+5.5}_{-1.8} \times 10^{22}$ \\
 $\Gamma$                        & $0.88^{+1.9}_{-0.85}$               & $1.5^{+1.2}_{-1.0}$ \\
$F_x$ (ergs~s$^{-1}$~cm$^{-2}$)\tablenotemark{a} & $4.8 \times10^{-13}$               & $4.2 \times10^{-13}$ \\
$\chi^2$~(dof)                   & \multicolumn{2}{c}{ 0.63(21) }
\enddata 
\tablenotetext{a}{Absorbed flux averaged for the two EPIC instruments in the 2$-$10~keV energy band.}
\label{spectab}
\end{deluxetable}

\subsection{Swift Observations} 

The TeV field of \tev\ was observed with the \swift\ X-ray telescope
(XRT) in photon counting mode on three occasions, logged in
Table~\ref{swifttab}. The first observation on 2005 June 29 UT is
described by \citet{lan06}. They identified three sources, of which two
are detected by \xmm\ at $>4\ \sigma$ significance. Two subsequent
observation, obtained on 2007 March 11 and 2008 February 15 UT, reveal a
total of seven sources detected at the $>3\ \sigma$ significance level
(marked with circles in Fig.~\ref{mosimg}). We extracted background
subtracted count rates for source~7 from each observation to search
for variability.  The counts derived from the third observation show a
$3\sigma$ increase over the first two, or $2\sigma$ over
the mean rate (see Table~\ref{swifttab}). This range of count rates
is consistent with the \xmm\ measured flux in
Table~\ref{spectab}. Using the {\tt PIMMS} software, the \xmm\
flux of source 7 predicts a 2$-$10~keV \swift\ count rate
of $(2-4) \times 10^{-3}$~s$^{-1}$.
Given the large uncertainty in both the \xmm\
spectral fit and the \swift\ counts, we can make no firm conclusion
about intrinsic flux variability of source 7, a possibility 
that cannot be excluded.


\section{Discussion} 

The spectral properties of source 7 distinguish it from the local
field stars by its hard, non-thermal emission. Coupled with the lack
of an optical/IR counterpart, this favors a compact object.  Its
spectrum is flatter than that of the typical AGN.  It is also notable
that source 7 lies at a plausible birth place for a pulsar, the center
of the shell-type SNR W41, and is located $3^\prime$ from the centroid
of \tev, well within its $10^\prime$ diameter ($1 \sigma$)
extent. Finally, its location at one of the vertexes of the patch of
diffuse X-ray emission uniquely suggests that source 7 is a pulsar
generating a PWN, \pwn, and also \tev. However, possible flux variability
among \swift\ pointings, if significant, would prefer 
an X-ray binary, unlikely to be associated
with the extended TeV source. Pending further evidence of variability, a pulsar
scenario remains the preferred hypotheses for the TeV emission.

In PWNe, relativistic $e^{\pm}$ are accelerated by the pulsar and
its wind termination shock, and TeV photons may be produced by inverse Compton
scattering of the cosmic microwave background, interstellar
and stellar IR/optical photons, or low-energy synchrotron photons.  If the
accelerated particles include protons, which collide with the ambient medium
making pions, TeV $\gamma$-rays may also be the product of $\pi^0$ decay.
For \tev, target material is readily available in the
GMC in the vicinity of the SNR W41 or in the supernova shell itself.
The {\it Midcourse Space Experiment} $21\mu$m survey data reveals
cold dust in the two lobes of the molecular cloud found in CO by
\citet{tia07}; however, these do
not align well with the diffuse X-ray emission. Finally, there are
several bright HII regions within $10^{\prime}$ of the TeV source that
are potential sources of seed photons.

\subsection{PWN X-ray and $\gamma$-ray Energetics} 

Energetically, if \psr\ is associated with W41 at a
distance of 4~kpc \citep{tia07}, its 2$-$10 keV luminosity is
$L_x=9\times 10^{32}\ d^2_{4}$ ergs~s$^{-1}$.  This implies a spin-down
luminosity from a rotation-powered pulsar of $\dot E \sim
10^{36}$~erg~s$^{-1}$ \citep{pos02}, near the empirical threshold
of $\dot E_c > 4\times 10^{36}$ ergs~s$^{-1}$ for generating a bright
($F_{PWN}/F_{PSR} \gtrsim 1$; 2$-$10~keV) PWN \citep{got04}. Our
measured flux ratio of $F_{PWN}/F_{PSR} \approx 1$ is at the low end
for pulsars above $\dot E_c$, but is similar to
some other \hess\ sources associated with PWNe. Most
notably, the 2$-$10 keV flux ratio for the energetic ($\dot E =
5.5\times 10^{36}$ ergs s$^{-1}$) pulsar PSR~J1838$-$0655 associated
with HESS~J1837$-$069 is $F_{PWN}/F_{PSR} \sim 0.1$ \citep{got08}.

\tev\ is detected up to 3~TeV. However, if we extrapolate
and integrate its $\Gamma=2.45$ power-law \citep{aha06a} from 0.3 to 30
TeV for comparison with \citet{gal08},
we get an energy flux of $1.4\times 10^{-11}$ ergs cm$^{-2}$
s$^{-1}$, corresponding to a luminosity of $L_{\gamma} = 2.7\times
10^{34}\ d^2_4$~ergs~s$^{-1}$ at the location of W41
\citep{alb06}. Under the assumption that the putative pulsar \psr\ powers the
observed TeV emission, the estimated efficiency ($0.3 - 30$~TeV) is
$\epsilon \equiv L_{\gamma}/\dot E \sim 3\%$. This is consistent
with the range of efficiencies ($0.01-11\%$) found for other
very-high-energy PWNe candidates \citep{gal08}.  Compared to
the X-ray luminosity, the $\gamma$-ray emission clearly dominates,
with a ratio of $L_{\gamma}/L_x \approx 29$.   This is similar to
the value $L_{\gamma}/L_x \approx 33$ found for the apparent
PWN association of G338.3$-$0.0 with HESS J1640$-$465 \citep{fun07}.
Several authors have hypothesized that such high flux ratios are
possible from inverse Compton scattering when PWNe are bathed
in infrared seed photons from nearby \ion{H}{2} regions, e.g.,
W33 close to HESS~J1813$-$178 \citep{hel07}, or the massive star
cluster RSGC1 near HESS~J1837$-$069 \citep{got08}.
A similar scenario may apply to the bright TeV PWNe that are
displaced from their pulsar power sources, e.g.,
HESS~J1825$-$137/PSR~B1823$-$13 \citep{pav08}.

\subsection{Alternative Hypotheses} 

\begin{deluxetable}{lccc}
\tablecolumns{4}
\tablewidth{0pt}
\tablecaption{\swift\ XRT Detections of \xmm\ Source 7}
\tablehead
{
Parameter  & 2005 Jun 29 & 2007 Mar 11 & 2008 Feb 15\\
           &     (UT)    &     (UT)    &   (UT)     
}
\startdata
ObsID                     & 00035159001 & 00035159002 & 00035117003 \\
Exposure (s)              & 5977        & 5019        & 6183\\
Counts\tablenotemark{a}   & $12\pm4.0$  & $9.8\pm3.6$ & $32\pm6.0$\\
Rate ($10^{-3}$~s$^{-1}$)\tablenotemark{a} & $2.0\pm0.7$ & $1.9\pm0.7$ & $5.1\pm1.0$

\enddata
\tablenotetext{a}{Backround subtracted counts and count rate
in a $1\farcm4$ diameter aperture in the 2$-$10~keV energy band.}
\label{swifttab}
\end{deluxetable}

One of the surprises of the TeV era is the offset of many
HESS sources from the PWNe nominally
responsible for powering the $\gamma$-ray emission.
With this displacement in
mind, some authors consider the 85~ms pulsar \apsr\ that
lies $24^{\prime}$ north of W41 as a potential source of accelerated
particles powering \tev.
A possible association of \apsr\
with W41 was mentioned by \citet{cli86} and was deemed plausible by
\citet{gae95}.  \citet{wei95} derived a kinematic distance of
4$-$5~kpc for \apsr\ from \ion{H}{1} absorption, which is compatible
with the \ion{H}{1} distance of $4.1\pm0.3$~kpc to W41 \citep{lea08}.
\citet{tia07} also noted that the $1.5\times 10^5$~yr characteristic
age of \apsr\ is consistent with the $\approx 10^5$~yr age that they
estimated for W41.  Furthermore, a large proper motion of $33 \pm 5$
mas~yr$^{-1}$ toward positive Galactic latitude, a direction away from
W41, was measured for \apsr\ by \citet{hob05}.  At $d=4.1$~kpc, this
corresponds to a not implausible tangential velocity $v_t = 640$
km~s$^{-1}$. In $1 \times 10^5$~yr, \apsr\ would have traveled
$55^{\prime}$, even further than its present offset from W41.
Although the spin-down luminosity
($\dot E = 5.8 \times 10^{35}$ ergs~s$^{-1}$)
of this pulsar could be sufficient to power \tev,
\citet{aha06a} felt that the large separation renders
an association unlikely.

Without regard to its known proper motion, \citet{bar08} advocate
\apsr\ as the source of \tev, assuming that the pulsar has
$v_t \sim 250$ km~s$^{-1}$.  They hypothesized that prior
injection of leptons by the pulsar while near its birth location
powers the currently observed TeV emission by inverse Compton scattering.
However, given our detection of a putative pulsar/PWN inside W41 and
coincident with \tev, we consider that our candidate
is more likely the source of \tev, probably with a spin-down luminosity 
$\dot E > 10^{36}$ ergs~s$^{-1}$, and that the nearby \apsr\ is
a chance coincidence, possibly a product of the frequent
birth of pulsars in active star formation regions. 
 
The location of GMCs near SNR W41 led 
\citet{tia07} to explore the possibility that the TeV emission from
\tev\ might result from the interaction between the two. \citet{yam06}
considered the production of TeV gamma rays in older SNR shells and
from the SNR shock running into a GMC, using the diagnostic
ratio of TeV to X-ray flux to distinguish between the different
possibilities.  They suggest that this ratio is of order 10 to 100 for
TeV gamma-ray emission from the SNR, but lower ($\sim 10$) for a SNR
shock running into a GMC.  The flux ratio $F_{\gamma}(1-10\ {\rm
TeV})/F_x(2-10\ \rm keV)$ obtained by \citet{tia07} and this work
is consistent with both of these scenarios. TeV gamma-ray
production due to the interaction of high energy protons from the SNR
shock with the GMC is, however, unlikely; based on \citet{yam06}, this
would require a much higher TeV/X-ray flux ratio of $> 100$.

\section{Conclusions}

The Galactic TeV
sources detected by HESS include, by our count, $\approx20$
PWN counterparts \citep[see also][]{gal08,hes08} and a few more possible
PWN associations, making this the largest class of
identified Galactic high energy sources.
While it is plausible that \tev\ is associated with the GMC located at
the center of W41 as argued by \citet{tia07}, we
believe that our detection of a putative pulsar/PWN candidate
\psr/\pwn\ within the 
extent of \tev\ suggests a more likely scenario.
The X-ray and $\gamma$-ray morphology and flux ratio
are consistent with the TeV source being powered by a pulsar/PWN.
Deeper X-ray timing observations can be performed
to search for pulsations from \psr,
which will be invaluable to test this scenario
and determine more quantitatively the energetics of this system.

\acknowledgements

RM was supported in part by the National Science Foundation
under grant no. PHY-0601112. EVG acknowledges {\sl Chandra} grant SAO AR7-8004X.

\end{document}